\documentclass[journal]{IEEEtran}
\IEEEoverridecommandlockouts
\usepackage{cite}
\usepackage{amsmath,amssymb,amsfonts}
\usepackage{algorithm}
\usepackage{algpseudocode}
\usepackage{graphicx}
\usepackage{adjustbox}

\graphicspath{ {figures/} }
\usepackage{textcomp}
\usepackage[dvipsnames]{xcolor}
\usepackage{tabularx}
\usepackage{multirow}
\usepackage{longtable}
\usepackage{booktabs}
\usepackage{soul}
\usepackage{arydshln}
\usepackage{threeparttable}
\usepackage{subcaption}
\PassOptionsToPackage{hyphens}{url}
\usepackage{hyperref}
\usepackage{authblk}

\begin{document}

\title{Emoception: Selective Affective Layer Fine-Tuning of Video Vision Transformers for Player Arousal Change Recognition from Gameplay Footage%
\thanks{\textcopyright~2026 IEEE. Personal use of this material is permitted. Permission from IEEE must be obtained for all other uses, in any current or future media, including reprinting/republishing this material for advertising or promotional purposes, creating new collective works, for resale or redistribution to servers or lists, or reuse of any copyrighted component of this work in other works. This work has been published in IEEE Transactions on Games. DOI: 10.1109/TG.2026.3691772}}

\author[1,*]{Yi Xia}
\author[2]{Ibrahim Khan}
\author[1]{Mury Fajar Dewantoro}
\author[3]{Wenwen Ouyang}
\author[2]{Ruck Thawonmas}

\affil[1]{Graduate School of Information Science and Engineering, Ritsumeikan University, Osaka, Japan\\
\texttt{yi.xia@ice.ci.ritsumei.ac.jp}, \texttt{fajar.dewantoro.mury@ice.ci.ritsumei.ac.jp}}
\affil[2]{College of Information Science and Engineering, Ritsumeikan University, Osaka, Japan\\
\texttt{khan@fc.ritsumei.ac.jp}, \texttt{ruck@is.ritsumei.ac.jp}}
\affil[3]{Alumni Association of Carnegie Mellon University, Pittsburgh, PA, USA\\
\texttt{wenweno@alumni.cmu.edu}}
\affil[*]{Corresponding author: \texttt{yi.xia@ice.ci.ritsumei.ac.jp}}

\maketitle

\begin{abstract}
This paper proposes Selective Affective Layer Fine-Tuning (SALFT), an efficient adaptation framework for Video Vision Transformers in player arousal recognition from gameplay. To bypass computationally expensive full fine-tuning, SALFT introduces a selection criterion based on the L2-norm change in layer parameters after brief adaptation, directly measuring representational shifts and providing a more stable basis than gradient-based alternatives. Evaluated via five-fold cross-validation on the Arousal Video Game AnnotatIoN dataset, SALFT achieves performance comparable to full fine-tuning across all games without statistically significant degradation ($p > 0.05$), while updating only $\approx$8\% of parameters (over 92\% reduction). Notably, in one game, SALFT consistently outperforms both full fine-tuning and the best baseline across all metrics and folds, reaching the theoretical minimum $p$-value ($p=0.0625$, exact two-sided Wilcoxon signed-rank test). Additionally, we introduce an interpretability method to trace attention patterns, enhancing model transparency. These results establish SALFT as an effective and efficient approach for affective game computing. GitHub repo for code, data, and supplementary materials: \url{https://github.com/Yi-Xia-2010/Emoception-SALFT}

\end{abstract}


\section{Introduction}
Recognizing players’ emotional dynamics during gameplay is an emerging focus in affective computing~\cite{10269152}. In particular, modeling arousal changes, a key dimension of affect, offers insight into player engagement and cognitive states. Recent studies have explored subject-agnostic arousal recognition in games using game footage video (GFV)\cite{makantasis2019pixels,9360471} or GFV combined with gameplay telemetry\cite{10.1007/978-3-031-86585-5_13}, without requiring physiological signals.

Building on this line of work, we target subject-agnostic arousal change recognition from GFV and propose Selective Affective Layer Fine-Tuning (SALFT), a layer-freezing fine-tuning framework for the Video Vision Transformer (ViViT)\cite{Arnab_2021_ICCV}. While ViViT effectively captures spatiotemporal patterns, its high complexity makes full fine-tuning computationally expensive and prone to overfitting with limited data. SALFT mitigates these issues by introducing a novel, task-oriented selection criterion, the L2-norm change in layer parameters, to identify and fine-tune only the most responsive transformer block, significantly improving parameter efficiency.

We evaluate SALFT on the Arousal Video Game Annotation (AGAIN) dataset~\cite{9816018}, which contains gameplay videos from nine games across three genres. Using rigorous 5-fold cross-validation with temporal block splitting and comprehensive statistical evaluation, we compare SALFT against multiple baselines including full ViViT fine-tuning, ResNet+LSTM, and exploratory LoRA implementations. Results show that SALFT achieves performance comparable to full fine-tuning on most games, while fine-tuning only a small fraction of parameters, and significantly outperforms it in specific cases (e.g., \texttt{Gun}). To complement these findings, we introduce an efficient interpretability method for ViViT that produces fine-grained spatiotemporal relevance maps, revealing how the model attends to gameplay dynamics when predicting arousal changes.

Emoception, a blend of emotion and inception, captures the conceptual foundation of our work: understanding how the player’s affective states, particularly arousal, emerge and evolve during gameplay. This reflects our focus on investigating temporally embedded emotional cues within game footage, as interpreted through the representational capacity of a spatiotemporal video model. Our contributions are as follows:
\begin{itemize}

    \item We propose SALFT, an efficient adaptation framework for ViViT that introduces a novel, directly measurable criterion (L2-norm change) for selecting the most task-responsive layer to fine-tune, improving both parameter efficiency and performance in arousal recognition.
    \item A rigorous 5-fold cross-validation evaluation on the AGAIN dataset demonstrates that SALFT consistently matches or numerically exceeds full fine-tuning performance across all nine games with only $\sim$8\% trainable parameters. Comparisons with strong baselines, including a pixel-based ResNet+LSTM, confirm its competitiveness.
    \item We propose an interpretability method for ViViT that generates efficient, fine-grained spatiotemporal relevance maps, offering insights into model decision-making for predicting arousal changes.
        
\end{itemize}

%%%%%%%%%%%%%%%%%%%%%%%%%%%%%%%%%%

\section{Related Work}

\subsection{Subject-Agnostic Affective Computing in Games}
A trend in affective game computing is the development of subject-agnostic models that infer player emotion from in-game signals alone, avoiding intrusive physiological sensors~\cite{10269152}. Seminal research using gameplay footage showed that deep networks can predict player arousal directly from raw game frames~\cite{makantasis2019pixels} and audio-visual streams~\cite{9360471}. More recently, large-scale annotated datasets such as AGAIN~\cite{9816018} have advanced gameplay telemetry analysis. Building on this, the study~\cite{10.1007/978-3-031-86585-5_13} demonstrates the potential of multimodal fusion through specialized architectures like ViViT that integrate video and context. Separately,~\cite{melhart2025can} showcases the power of Multimodal Large Language Models to process visual and text prompts for inferring player engagement. These studies highlight video’s dual role as both an affective cue and a stimulus, motivating our focus on efficient strategies for subject-agnostic arousal recognition from gameplay footage.

\subsection{Efficient Adaptation of Video Transformers}
Transfer learning with large pre-trained video transformers such as ViViT~\cite{Arnab_2021_ICCV} has proven effective for downstream tasks like player arousal recognition~\cite{10.1007/978-3-031-86585-5_13}, but full fine-tuning remains computationally expensive and susceptible to overfitting. To mitigate this, strategies such as explicit regularization~\cite{xuhong2018explicit, gouk2021distance, zhou2023dr} and Parameter-Efficient Fine-Tuning (PEFT) have been proposed. PEFT methods reduce trainable parameters through various means, including addition-based techniques like Adapters~\cite{pmlr-v97-houlsby19a} and Soft Prompts~\cite{li-liang-2021-prefix, liu-etal-2022-p}, reparameterization approaches such as LoRA~\cite{hu2022lora} and its variants~\cite{liu-etal-2024-aflora, ding2023sparse}, and selection-based methods that fine-tune a subset of parameters~\cite{guo-etal-2021-parameter, howard-ruder-2018-universal, lee2023surgical}.

Recent selective fine-tuning approaches face practical limitations. SubTuning requires conducting multiple experiments to construct a fine-tuning profile for block selection, where each experiment fine-tunes only one block while freezing the others; this process is computationally intensive~\cite{kaplun2023less}. AIM-Fair relies on synthetic data generation~\cite{zhao2025aim}. Feedforward-based Parameter Selection (FPS) ranks parameters by the product of magnitudes and input activations~\cite{yang2025fps}. While FPS efficiently identifies salient parameters through a deterministic, forward-pass-only metric, its selection is static and decoupled from the optimization dynamics that drive adaptation during fine-tuning. Furthermore, the lack of an official implementation precludes rigorous reproducibility.

Most closely related to our work is surgical fine-tuning~\cite{lee2023surgical}, which efficiently adapts a single contiguous block selected via heuristic criteria like Relative Gradient Norm (RGN) or Signal-to-Noise Ratio (SNR). However, as shown in the original work, these heuristics do not consistently select the block that maximizes final accuracy compared to an exhaustive per-block validation strategy (``Cross-Val''), revealing a reliability gap. Motivated by the need for a more direct and stable selection signal than the optimization-sensitive gradient heuristics, we propose SALFT. Our key innovation is a task-oriented selection heuristic: the L2-norm change in layer parameters after brief fine-tuning. This metric directly quantifies how much a layer's parameters adjust in response to the affective recognition task, offering a more stable and interpretable criterion for layer responsiveness. Our approach enhances arousal recognition while maintaining the efficiency of single-run, data-parsimonious adaptation.

\subsection{Explainability in Video Transformers}

While visual attribution methods were largely pioneered for CNNs, evolving from the seminal Class Activation Mapping (CAM)~\cite{zhou2016learning} to influential successors like Grad-CAM~\cite{selvaraju2017grad} and others~\cite{chattopadhay2018grad, 9462463, belharbi2025colocam}, their architectural specificity necessitated the development of new approaches tailored for Transformers. For Transformers, these methods range from class-agnostic attention visualization like Rollout~\cite{abnar-zuidema-2020-quantifying} to class-specific attribution. The latter typically either propagate relevance backward through the network, as in the Layer-wise Relevance Propagation (LRP)-based work of Chefer et al.\cite{Chefer_2021_CVPR}, or fuse attention with gradients at the layer level, as in Grad-SAM\cite{10.1145/3459637.3482126}. Complementing these attribution methods, other works like Video Transformer Concept Discovery~\cite{kowal2024understanding} focus on discovering higher-level semantic concepts.

Recognizing player arousal from subtle, transient gameplay cues requires fine-grained spatiotemporal attribution, a need unmet by concept-based methods like VTCD~\cite{kowal2024understanding}. To this end, we propose a hybrid method for ViViT that integrates the gradient-weighting principle of Grad-SAM~\cite{10.1145/3459637.3482126} with the multi-layer rollout framework of Chefer et al.~\cite{Chefer_2021_CVPR}. This design mitigates the prohibitive computational cost of directly adapting LRP to ViViT while still producing pixel-level spatiotemporal maps for detailed attribution.

%%%%%%%%%%%%%%%%%%%%%%%%%%%%%%%%%

\section{Methodology}

\subsection{Task Reformulation and data preprocessing}
In the original AGAIN dataset study~\cite{9816018}, arousal change prediction was formulated as a binary classification task. Training samples were generated by applying a threshold to the magnitude of arousal change, retaining only segments with significant increases or decreases while discarding stable periods. The resulting model was trained to predict whether arousal would increase or decrease. Although this formulation represented a significant step toward modeling arousal changes in game contexts, it limits applicability to real-time scenarios where arousal may remain stable.

To better reflect real-world scenarios, the work in~\cite{10.1007/978-3-031-86585-5_13} reformulated the task as a three-class classification problem, predicting whether arousal would increase, decrease, or remain stable. Unlike the aforementioned threshold-based approach, change direction was determined by directly comparing the mean arousal values of consecutive windows. Furthermore, 1-second sliding windows were employed to expand the dataset and capture fine-grained temporal dynamics. 

Following~\cite{10.1007/978-3-031-86585-5_13}, we adopt the same preprocessing strategy while focusing exclusively on video data. For each time point $t$, a sample consists of two consecutive 3-second windows ($t$ to $t+6$), yielding a 6-second segment. All samples are retained regardless of change magnitude. For each segment, 32 uniformly spaced frames are extracted, resized to 224×224 pixels, and fed into the ViViT-based modeling pipeline to match its input specification. Threshold selection and tuning, as conducted in~\cite{9360471,9816018}, are beyond the scope of this study and thus not considered. Following preprocessing, the number of samples per arousal change class across all games is shown in the last three columns of Table~\ref{tab:game_list}.

\subsection{Selective Affective Layer Fine-Tuning}\label{method:exploratory}
Our SALFT framework consists of two consecutive stages: (1) \textbf{exploratory full fine-tuning} to identify the layer whose parameters deviate most from the pretrained initialization, and (2) \textbf{second-stage selective fine-tuning} in which only the identified layer remains trainable while all others are frozen. Unlike Surgical Fine-Tuning which relies on indirect, gradient-based heuristics (RGN/SNR) to approximate layer importance, our criterion directly measures the actual parameter change induced by the task. This provides a more stable and interpretable signal for selecting the most responsive layer. Throughout the entire fine-tuning framework, the standard cross-entropy loss is consistently employed for model optimization. The procedure is illustrated in Algorithm~\ref{alg:selective-finetuning} and detailed below.

\begin{algorithm}[tbp]
\caption{Selective Affective Layer Fine-Tuning}
\label{alg:selective-finetuning}
\begin{algorithmic}[1]
\Require Pretrained parameters $\Theta^{(0)}$; correlation threshold $\tau_\rho = 0.80$
\Ensure Final model with only one non-classifier layer and the classifier layer fine-tuned 
\Statex \textit{//First-Stage: Exploratory Full Fine-Tuning} 
\State Initialize epoch counter: $e \gets 0$
\State Initialize layer-wise distance vector: $\boldsymbol{\Delta}^{(0)} \gets \mathbf{0}$
\Repeat
    \State $e \gets e + 1$
    \State Fine-tune all layers for one epoch using current training data
    \For{$l = 1$ to $L-1$}
        \State $\mathbf{v}^{(e)}_l \gets \textsc{ConcatFlatten}(\boldsymbol{\theta}^{(e)}_{l,*})$
        \State $\Delta^{(e)}_l \gets \|\mathbf{v}^{(e)}_l - \mathbf{v}^{(0)}_l\|_2$
    \EndFor
    \State $\boldsymbol{\Delta}^{(e)} \gets [\Delta^{(e)}_1, \dots, \Delta^{(e)}_{L-1}]$
    \State $(\rho, p) \gets \textsc{Spearman}(\boldsymbol{\Delta}^{(e-1)}, \boldsymbol{\Delta}^{(e)})$
\Until{$\rho \ge \tau_\rho$ \textbf{and} $p < 0.05$}

\State $e^\star \gets e$
\State $l^\star \gets \arg\max_l \Delta^{(e^\star)}_l$

\Statex \textit{//Second-Stage: Selective Fine-Tuning}
\State Reset model parameters to $\Theta^{(0)}$
\State Freeze all layers except $l^\star$ and $L$ 

\While{not converged}
    \State Fine-tune only layers $l^\star$ and $L$
\EndWhile

\Return The selectively fine-tuned model
\end{algorithmic}
\end{algorithm}

\textit{First-Stage: Exploratory Full Fine-Tuning:}

\begin{enumerate}
    
    \item \textbf{Initial Setup.} 
    Let \(\Theta^{(0)} = \{\boldsymbol{\theta}^{(0)}_1, \dots, \boldsymbol{\theta}^{(0)}_L\}\) denote the pretrained parameters of all layers \(l \in \{1, \dots, L\}\), where each \(\boldsymbol{\theta}_l\) includes both weights and biases. We fine-tune the entire model on the downstream corpus, saving a checkpoint at the end of each epoch \(e\). Fine-tuning continues until the pattern of layer-wise parameter changes is deemed stable according to a correlation-based convergence criterion. Since the classifier head, is re-initialized to match the downstream class set, its parameters are expected to change accordingly. Consequently, we exclude the classifier layer, $L$, from the displacement computations and stability analysis; all subsequent operations focus exclusively on the remaining \(L-1\) non-classifier layers.   
    
    \item \textbf{Layer-wise parameter aggregation.}  
    For each non-classifier layer \(l\), which may comprise multiple sub-components (e.g., query, key, value projections and feed-forward blocks), we aggregate the parameters at epoch~\(e\) by flattening and concatenating all parameters within the layer into a single vector:
    \begin{equation}
    \mathbf{v}^{(e)}_l
      = \operatorname{concat}\!\bigl(
          \textsf{vec}(\mathbf{\boldsymbol{\theta}}^{(e)}_{l,1}),\;
          \dots,\;
          \textsf{vec}(\mathbf{\boldsymbol{\theta}}^{(e)}_{l,K_l})
        \bigr)
      \in \mathbb{R}^{d_l},
    \end{equation}
    where \(K_l\) is the number of sub-modules in layer~\(l\) and \(d_l\) the resulting dimensionality.
    
    \item \textbf{Layer-wise displacement.}  
    For each layer, we compute the \(\ell_2\) distance between its parameter vector at epoch \(e\) and the corresponding original parameter vector, defined as
    \begin{equation}
        \Delta^{(e)}_l = \left\| \mathbf{v}^{(e)}_l - \mathbf{v}^{(0)}_l \right\|_2,
    \end{equation}
    where \(\mathbf{v}^{(e)}_l\) denotes the parameter vector of layer \(l\) at epoch \(e\), and \(\mathbf{v}^{(0)}_l\) denotes the original pretrained parameter vector of the same layer.
    We then form a displacement vector using the displacement values (2) as follows:
    \begin{equation}
    \boldsymbol{\Delta}^{(e)} = \bigl(\Delta_1^{(e)}, \Delta_2^{(e)}, \ldots, \Delta_{L-1}^{(e)}\bigr).
    \end{equation}

    \item \textbf{Stability check.} To assess the stabilization of the layer-wise displacement pattern, we compute the Spearman rank correlation coefficient \(\rho^{(e)}\) between \(\boldsymbol{\Delta}^{(e-1)}\) and \(\boldsymbol{\Delta}^{(e)}\). We consider the pattern to be stabilized when the correlation is both statistically significant and exceeds a predefined threshold \(\tau_\rho\), as specified by the following condition:
    \begin{equation}
    \rho^{(e)} \ge \tau_\rho = 0.80 \quad \text{and} \quad p\text{-value} < 0.05.
    \label{eq:stabilization_criterion}
    \end{equation}
    This criterion indicates that the trend of displacements across layers remains consistent between successive epochs. The threshold \(\tau_\rho\) is motivated by commonly accepted interpretations of Spearman correlation coefficients, where values above 0.70 or 0.80 are generally regarded as reflecting strong or stable associations~\cite{evans1996straightforward, mukaka2012guide, schober2018correlation}. Note that this criterion is used solely to assess the stability of ranking trends; it does not imply linearity or causation. We define \(e^\star = e\) as the last epoch before this condition is first met, and terminate fine-tuning at epoch \(e^\star\).
    
    \item \textbf{Layer selection.} The layer with the largest displacement at epoch \( e^\star \) is selected:
    \begin{equation}
    l^\star = \arg\max_{l} \Delta^{(e^\star)}_l.
    \end{equation}

\end{enumerate}

\textit{Second-Stage: Selective Fine-Tuning:}
We re-initialize the model to \( \Theta^{(0)} \), then perform a second fine-tuning run in which only the selected layer \( l^\star \) and classifier head $L$ are trainable; all other layers are frozen. The optimizer state is reset to avoid interference from the previous fine-tuning stage. This strategy concentrates the model’s learning capacity on the most task-relevant layer while preserving the generalization ability of the pretrained model.

\subsection{Model interpretability}\label{method:interpretation}

To interpret the spatiotemporal decision-making of ViViT models, we propose an attribution method inspired by the approaches of Chefer et al.~\cite{Chefer_2021_CVPR} and Grad-SAM~\cite{10.1145/3459637.3482126}. Chefer et al.'s approach computes layer-wise relevance via a backward propagation process using the computationally intensive LRP algorithm, which is then combined with class-specific gradients and aggregated across multiple layers to produce refined attributions. Grad-SAM combines gradients with attention and simplifies multi-layer attribution by averaging the attribution maps across layers. Our method aims to balance effectiveness and computational efficiency by integrating these ideas.

Our work, therefore, synthesizes these insights: we replace the expensive LRP relevance with the model's raw attention maps $\mathbf{A}$, integrating them into the sophisticated, non-averaging aggregation framework of Chefer et al. and adapting this process for ViViT's spatiotemporal architecture. This results in a lightweight signal that maintains high explanatory fidelity at a fraction of the computational cost.

In the following, we detail our attribution method adapted to the ViViT architecture. For each Transformer encoder layer $l$, we extract the raw multi-head self-attention maps $\mathbf{A}^l \in \mathbb{R}^{H \times N \times N}$, where $H$ is the number of heads and $N$ the number of tokens. Averaging over heads yields
\[
\bar{\mathbf{A}}^l = \frac{1}{H} \sum_{h=1}^{H} \mathbf{A}^l_h \in \mathbb{R}^{N \times N}.
\]
A backward pass is initialized from the target class logit $y_c$ to compute the class-specific gradients 
$\frac{\partial y_c}{\partial \mathbf{A}^l}$ for each layer $l$, which are head-averaged to obtain $\bar{\mathbf{G}}^l$. 

We define the layer-wise attention relevance matrix at layer $l$ as
\begin{equation}
    \mathbf{R}^l = \mathrm{ReLU}\big(\bar{\mathbf{A}}^l \odot \bar{\mathbf{G}}^l\big) + \mathbf{I},
\end{equation}
where $\odot$ denotes the Hadamard product and identity matrix $\mathbf{I}$ accounts for residual connections. 
Note that $\mathbf{R}^l$ here is an intermediate relevance matrix capturing gradient-weighted attention patterns within layer $l$, which is distinct from the relevance computed via LRP defined in~\cite{Chefer_2021_CVPR}. To ensure numerical stability and preserve token-wise relevance, each $\mathbf{R}^l$ is row-normalized so that each row sums to one before recursive aggregation. Starting from $\mathbf{C}^L=\mathbf{I}$ at the output layer $L$, cumulative relevance matrix is propagated backward via
\begin{equation}
    \mathbf{C}^{l-1} = \mathbf{C}^l \cdot \mathrm{normalize}(\mathbf{R}^l), \quad l = L, \dots, 1.
\end{equation}
The final relevance matrix $\mathbf{C}^0$ encodes the token-level relevance scores with respect to the target class, serving as the basis for the model’s explanation. 

Unlike 2D Vision Transformers, ViViT operates on spatiotemporal tubelets, where each token corresponds to a 2D spatial patch extended over a fixed number of consecutive frames. We derive the spatiotemporal relevance map $\mathbf{M}_{\text{rel}} \in \mathbb{R}^{T \times H \times W}$ by selecting the \texttt{[CLS]} row from $\mathbf{C}^0$ and reshaping it into a spatiotemporal grid, with $T$ denoting the number of tubelets and $H \times W$ the spatial token layout. Each temporal slice is then replicated over its corresponding frames to produce per-frame relevance heatmaps, which are upsampled to $224 \times 224$ and overlaid on the corresponding input video frames, providing a fine-grained visualization of the model’s spatiotemporal attribution.

%%%%%%%%%%%%%%%%%%%%%%%%%%%%%%%

\section{Experiment}\label{sec:experiment}

\subsection{Dataset}\label{sec:dataset}

\begin{figure}[tbp] 
    \centering

    \begin{subfigure}[b]{0.32\linewidth}
        \centering
        \includegraphics[width=\linewidth]{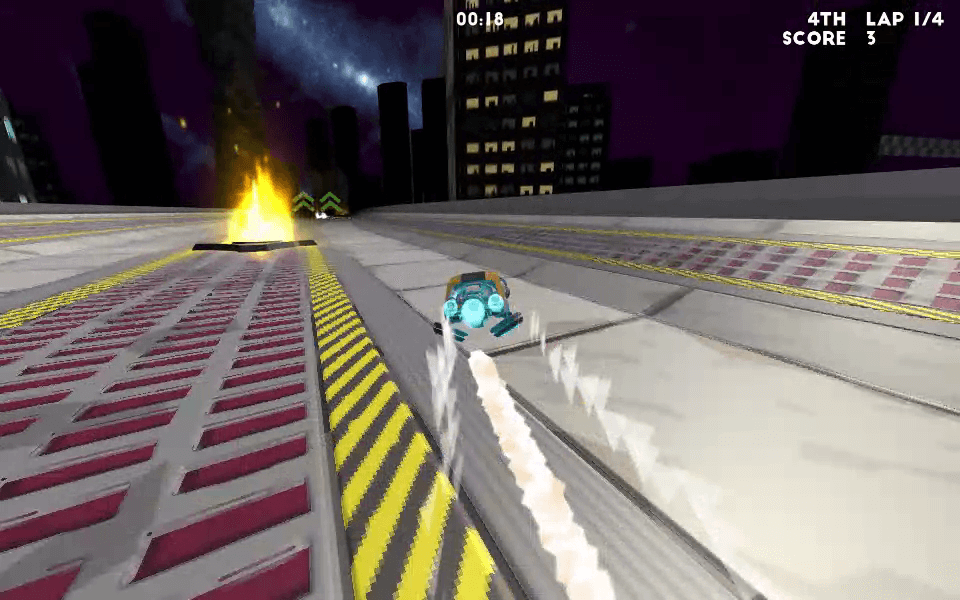}
        \caption{\texttt{apex}}
        \label{fig:apex}
    \end{subfigure}
    \begin{subfigure}[b]{0.32\linewidth}
        \centering
        \includegraphics[width=\linewidth]{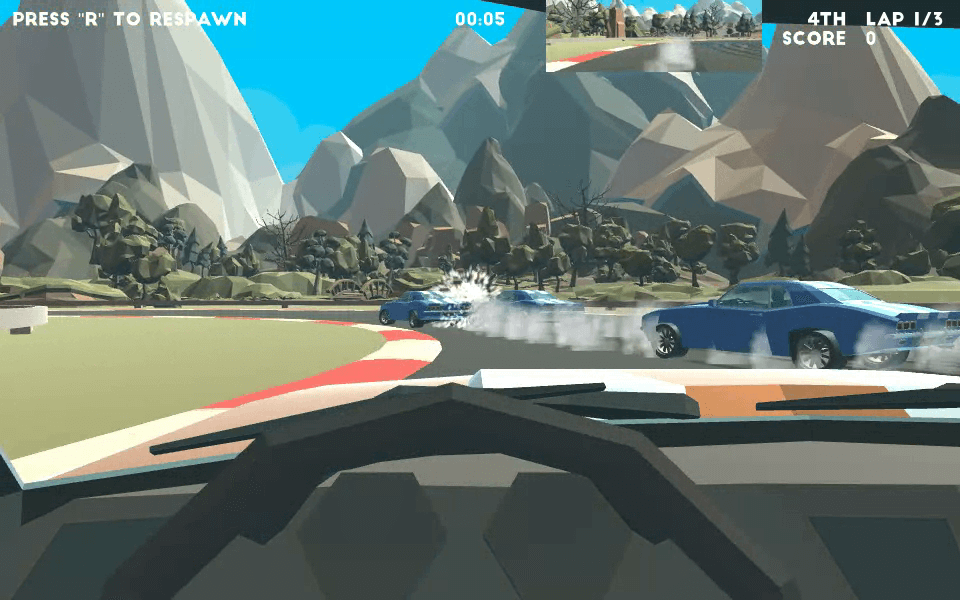}
        \caption{\texttt{solid}}
        \label{fig:game2}
    \end{subfigure}
    \begin{subfigure}[b]{0.32\linewidth}
        \centering
        \includegraphics[width=\linewidth]{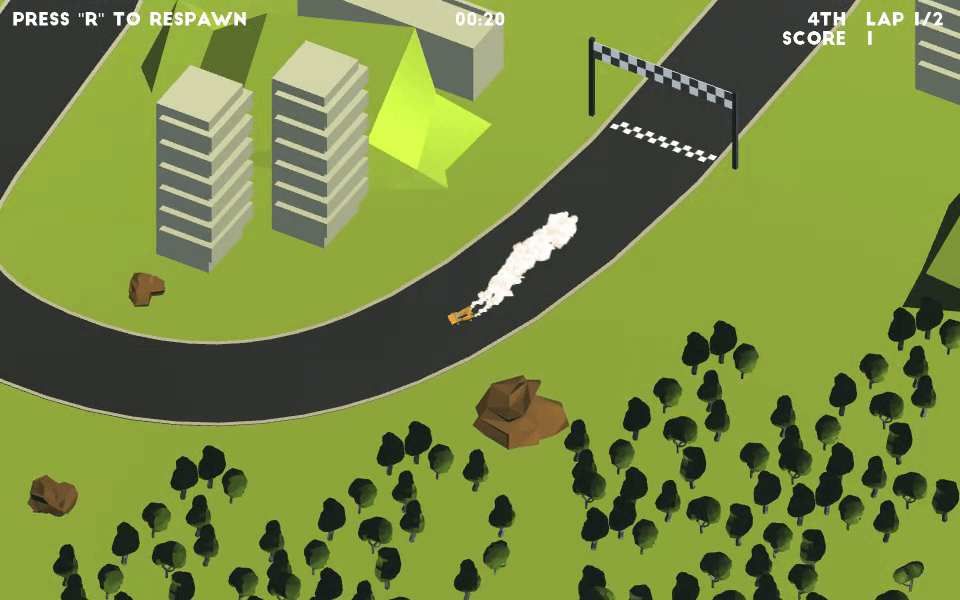}
        \caption{\texttt{tiny}}
        \label{fig:game3}
    \end{subfigure}

    \begin{subfigure}[b]{0.32\linewidth}
        \centering
        \includegraphics[width=\linewidth]{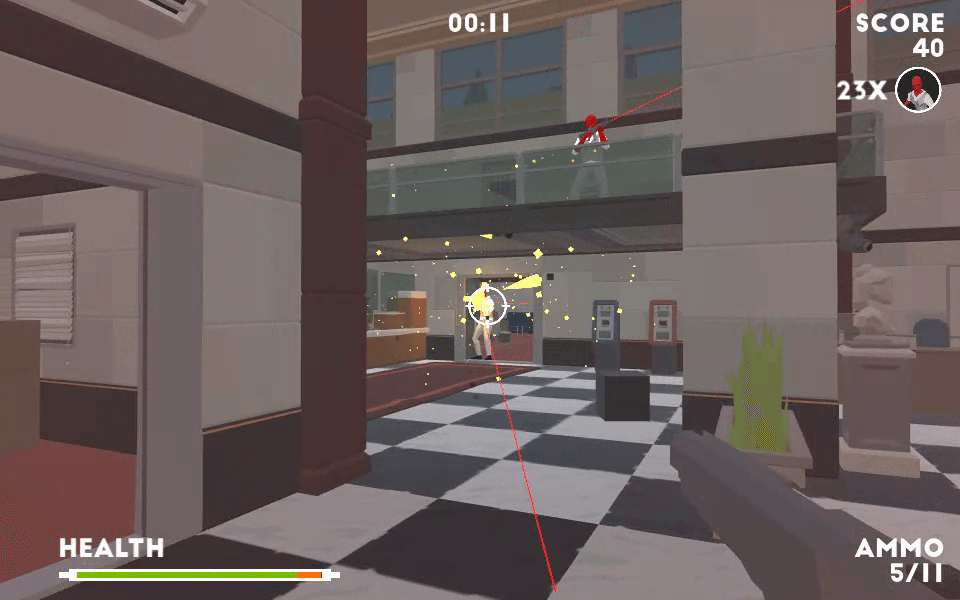}
        \caption{\texttt{fps}}
        \label{fig:game4}
    \end{subfigure}
    \begin{subfigure}[b]{0.32\linewidth}
        \centering
        \includegraphics[width=\linewidth]{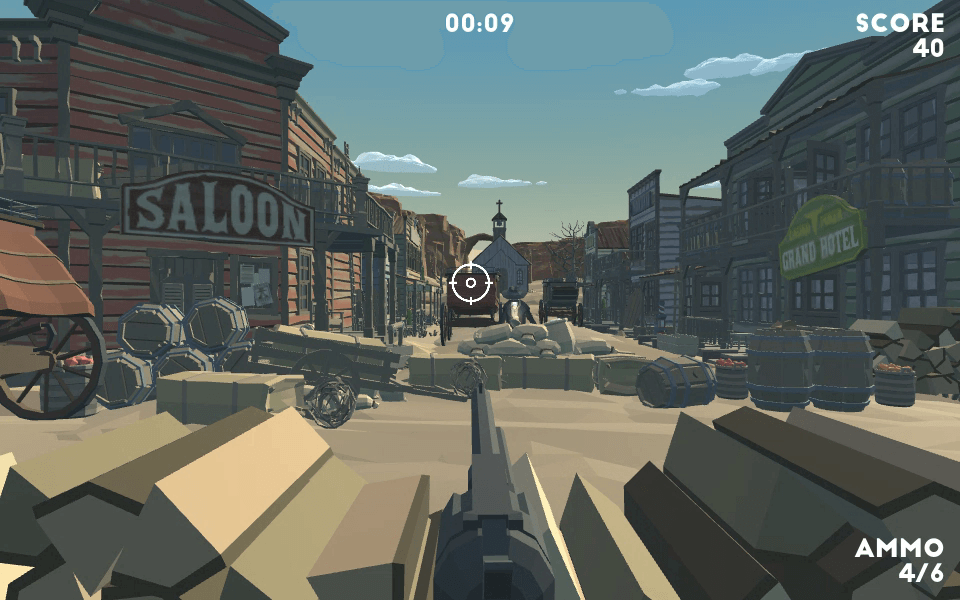}
        \caption{\texttt{gallery}}
        \label{fig:game5}
    \end{subfigure}
    \begin{subfigure}[b]{0.32\linewidth}
        \centering
        \includegraphics[width=\linewidth]{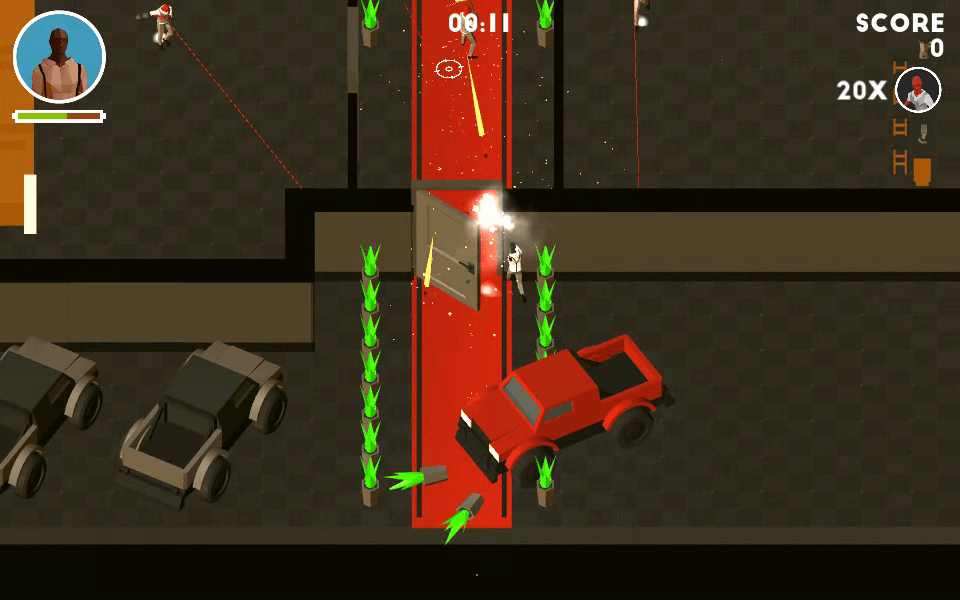}
        \caption{\texttt{topdown}}
        \label{fig:game6}
    \end{subfigure}

    \begin{subfigure}[b]{0.32\linewidth}
        \centering
        \includegraphics[width=\linewidth]{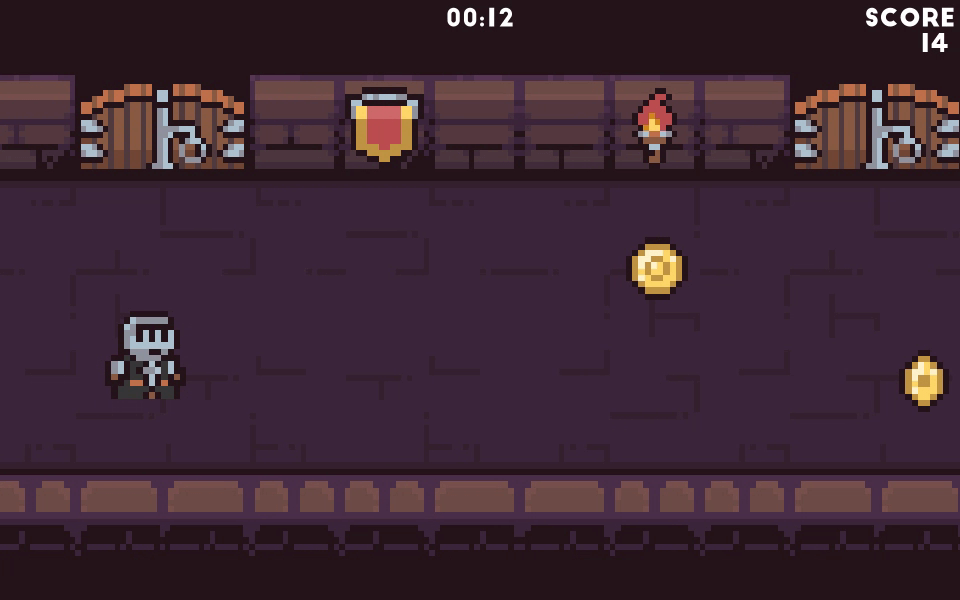}
        \caption{\texttt{endless}}
        \label{fig:game7}
    \end{subfigure}
    \begin{subfigure}[b]{0.32\linewidth}
        \centering
        \includegraphics[width=\linewidth]{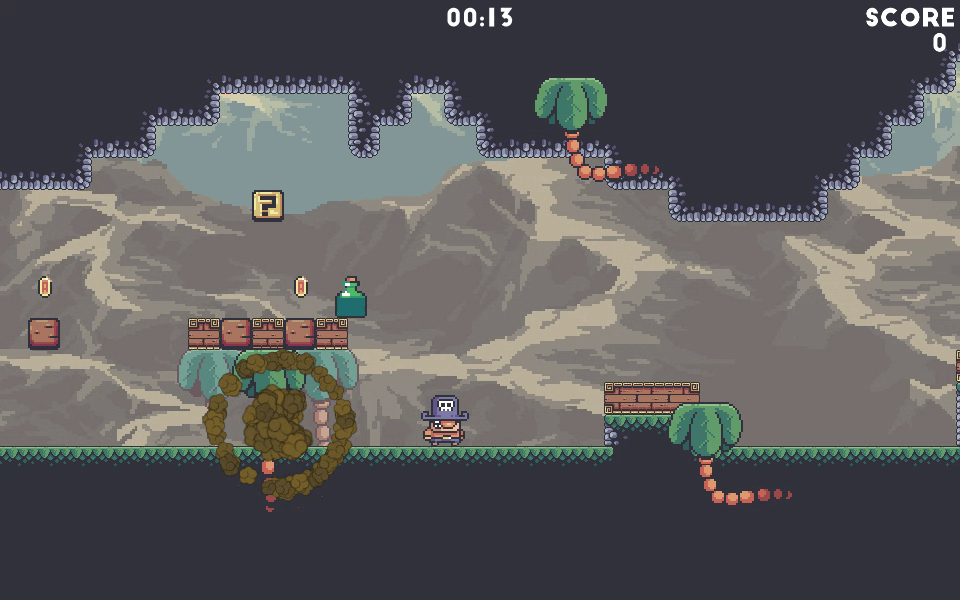}
        \caption{\texttt{platform}}
        \label{fig:game8}
    \end{subfigure}
    \begin{subfigure}[b]{0.32\linewidth}
        \centering
        \includegraphics[width=\linewidth]{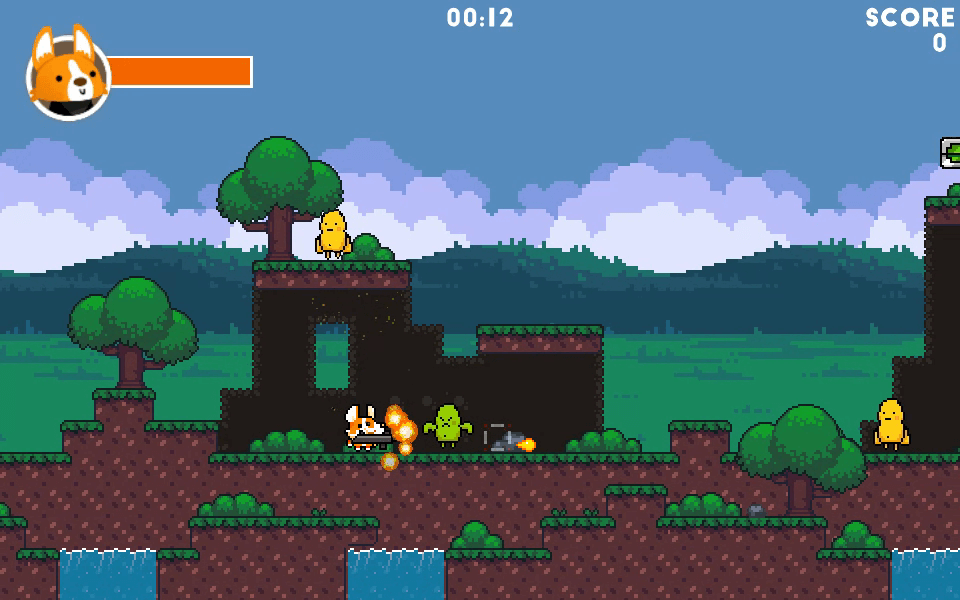}
        \caption{\texttt{gun}}
        \label{fig:game9}
    \end{subfigure}

    \caption{Representative screenshots of each game in the AGAIN dataset.}
    \label{fig:games_collection}
\end{figure}

This study employs the `clean' subset of the AGAIN dataset~\cite{9816018}, which comprises 995 two-minute gameplay sessions (from an original total of 1,116) from 122 participants. Each session consists of video recordings, in-game telemetry, and time-continuous self-reported arousal annotations. Data were collected from nine custom-developed games spanning racing, shooter, and platformer genres (Figure~\ref{fig:games_collection}), using the abbreviated identifiers provided in the dataset files (the first three columns of Table~\ref{tab:game_list}). The annotations were initially unbounded and subsequently normalized to the [0,1] range. Unlike previous approaches combining video and telemetry modalities~\cite{10.1007/978-3-031-86585-5_13}, this work focuses exclusively on video and its associated arousal traces, given that the visual modality provides more informative cues for this task.

\begin{table}[tbp]
\caption{Games in the AGAIN dataset with number of samples per arousal change class. The column ``Abbrev.'' stands for the abbreviated game identifier; 
``Inc.'', ``Dec.'', and ``Unchg.'' denote arousal Increase, Decrease, and Unchanging, respectively.}
\label{tab:game_list}
\centering
\begin{tabular}{lllr r r}
\toprule
\textbf{Game} & 
\textbf{Abbrev.} & 
\textbf{Genre} &
\textbf{Inc.} &
\textbf{Dec.} &
\textbf{Unchg.} \\
\hline
ApexSpeed     & \texttt{apex}     & Racing     & 6629 &  4446 &  2375 \\
Solid         & \texttt{solid}    & Racing     &  6520 &  4333 &  1947 \\
TinyCars      & \texttt{tiny}     & Racing     & 5944 &  4402 &  2279 \\
Heist         & \texttt{fps}      & Shooter    & 6561 &  3986 &  2337 \\
Shootout      & \texttt{gallery}  & Shooter    & 7049 &  3121 &  2236 \\
TopDown       & \texttt{topdown}  & Shooter    & 6914 &  3963 & 2724 \\
The Endless   & \texttt{endless}  & Platformer &  6610 &  3591 &  3010 \\
Pirates       & \texttt{platform} & Platformer &  6156 &  3748 &  2593 \\
Run`N'Gun     & \texttt{gun}      & Platformer & 6438 &  3863 &  2855 \\
\bottomrule
\end{tabular}
\end{table}

\subsection{ Data Split}
To ensure rigorous evaluation while addressing the specific spatiotemporal characteristics of gameplay data, we employed a Block-Stratified 5-Fold Cross-Validation strategy. Our design rationale is threefold:
\begin{itemize}
    \item \textbf{Addressing Interaction-Driven Covariate Shift:} Unlike passive media consumption, gameplay footage exhibits substantial variance due to diverse player strategies (e.g., aggressive vs. stealthy playstyles). A strict subject-level split risks evaluating the model on unseen interaction patterns, potentially conflating distinct playstyles with affect labels. By distributing user data across folds, we expose the model to the full manifold of gameplay behaviors, encouraging the learning of generalizable visual-state mappings rather than overfitting to specific interaction subsets.
    \item \textbf{Validating Affect-Content Independence:} A primary concern in random splitting is identity leakage. However, our analysis demonstrates that \textit{User Identity} is a poor predictor of affect in this domain. As shown in the GitHub repo, t-SNE visualizations reveal that feature embeddings from different players are highly intermixed, indicating no distinct ``user signatures'' in the gameplay video. Furthermore, the high Intra-User Label Entropy (mean=0.75) confirms that individual sessions contain a diverse spectrum of arousal states. This high within-subject variance forces the model to learn content-based features rather than relying on trivial subject-specific cues.
    \item \textbf{Preventing Short-Term Data Leakage:} We extract samples using a 1-second sliding window. Random splitting would introduce severe data leakage due to the high overlap between adjacent windows. To rigorously isolate this risk, we introduced non-overlapping temporal blocks as the atomic grouping unit. This ensures that explicit data overlap from the sliding window process is eliminated, maintaining strict disjointness between training and test sets at the sample level.
\end{itemize}

Our data partitioning protocol consists of three stages designed to prevent temporal leakage and maintain class balance. First, temporal micro-blocking segments each gameplay session into non-overlapping 6-second blocks, with all 1-second sliding windows within a block sharing a group identifier. This ensures that samples from the same local temporal context remain together. Second, a stratified 5-fold cross-validation scheme is applied at the block level, preserving the distribution of arousal classes in each fold while keeping all windows from a given block within a single fold. This eliminates intra-block leakage and disrupts session-level temporal continuity. Finally, a nested split reserves 20\% of blocks for testing, and divides the remaining 80\% into training (72\%) and validation (8\%) sets, maintaining the same block-level grouping constraints. This protocol rigorously isolates temporal dependencies while providing a balanced evaluation across affective states.

\subsection{Experimental Setup}

We adopt the ViViT-B/16$\times$2 backbone~\cite{Arnab_2021_ICCV}, pretrained on Kinetics-400, and replace the classification head to match the target classes while keeping the backbone unchanged. Experiments are conducted on the AGAIN dataset comprising nine games (Section~\ref{sec:dataset}), where the model is independently fine-tuned and evaluated on each game to assess generalization across diverse gameplay scenarios. Fine-tuning is performed on an NVIDIA A100 GPU using the AdamW optimizer with a batch size of 4 and a learning rate of $5\times10^{-5}$; other optimizer parameters are kept at their defaults. All results are reported from a single run due to computational constraints, with a fixed random seed (42) to ensure reproducibility. Spearman’s rank correlations are computed using SciPy’s \texttt{spearmanr}, with significance assessed at $\alpha=0.05$.  Code, data, and supplementary materials are publicly available (see abstract).

\subsection{Baseline and Comparison Strategy}\label{exp:comparison}
To evaluate the proposed fine-tuning framework, we compare it against several representative baselines. For a fair comparison, all methods are fine-tuned for 14 epochs, and we report the test performance achieved by the model checkpoint with the best validation performance. The 14-epoch limit is derived from the gradual unfreezing schedule~\cite{howard-ruder-2018-universal}, which we adapted to align with the ViViT model’s 14-layer architecture (embedding, 12 encoder layers, and classifier); however, following prior work reporting its inferior performance\cite{lee2023surgical}, this schedule was excluded as a baseline.

Our baseline selection also reflects practical computational constraints. For instance, while the Surgical Fine-Tuning paper~\cite{lee2023surgical} employs a rigorous cross-validation protocol (``Cross-Val'') for selecting the optimal layer block, it requires training a separate model for each candidate block, a cost-prohibitive for our large-scale video model across nine domains. This inefficiency directly motivates SALFT's one-pass selection protocol. The reliability of this efficient approach is supported by a preliminary study (c.f. the GitHub repo), which shows that our L2 criterion selects the same layer as the Cross-Val.

\textbf{Random Forest.} Following the original AGAIN dataset paper~\cite{9816018}, we implement a random forest model trained on game content features. Although our task differs by focusing on arousal changes prediction from GFV, this classical baseline provides a reference point. Prior work~\cite{10.1007/978-3-031-86585-5_13} reported a performance decline under the new task but evaluated only a single game; here, we extend the analysis to all nine games.

\textbf{ViViT Full Fine-tuning.} As a strong baseline, we fine-tune all parameters of the pretrained ViViT model. This standard approach in video understanding often achieves competitive results due to full model adaptability. Previous work~\cite{10.1007/978-3-031-86585-5_13} applied this to GFV, yielding strong performance for arousal change recognition. However, it incurs higher computational cost and risk of overfitting, especially with small or domain-shifted datasets.

\textbf{ResNet+LSTM (Full Fine-tuning).} Motivated by~\cite{10.1145/3577190.3614203}, to benchmark against a different and widely used architecture for spatiotemporal modeling, we include a 3D-ResNet-50+LSTM model, which is fully fine-tuned on our task. This baseline allows us to evaluate the relative performance of the ViViT backbone and our method within a different architectural family.

\textbf{SALFT on ResNet+LSTM.} To demonstrate the generality of our proposed SALFT framework beyond the ViViT architecture, we also apply its two-stage procedure (exploratory full tuning followed by selective layer tuning) to the ResNet+LSTM model. This tests whether the task-aware layer selection criterion is effective and beneficial for a substantially different network structure.

\textbf{ViViT with LoRA.} To compare against a modern parameter-efficient fine-tuning (PEFT) method, we implement Low-Rank Adaptation (LoRA)~\cite{hu2022lora} on the ViViT backbone. We configure LoRA to adapt the query, key, value, and final dense projection layers within each Transformer block. Starting from a standard configuration (learning rate \(5\times10^{-4}\), rank \(r=16\), scaling factor \(\alpha=32\)), we observed that LoRA failed to converge (predicting only a single class). Even after increasing the learning rate to \(1\times10^{-3}\), rank to \(r=64\), and \(\alpha\) to 128, the model still did not converge. This baseline is included to illustrate the challenge of applying generic PEFT methods to our gameplay video task, and thereby highlights the advantage of SALFT's task-aware, layer-wise selection strategy which successfully adapts the model.

\subsection{Evaluation.} 
Model performance is evaluated using three metrics: the weighted-average $F_1$-score (primary), the macro-average $F_1$-score, and accuracy. The weighted $F_1$-score is prioritized as it accounts for class frequency, providing a performance measure that better reflects the imbalanced label distribution. The macro $F_1$-score complements this by treating all classes equally, highlighting performance on minority classes. Accuracy is reported for completeness and direct comparison with prior work.

All results are reported as the mean and its 95\% confidence interval across the five cross-validation folds. Performance differences are assessed via exact two-sided Wilcoxon signed-rank tests separately for each game and metric, comparing ViViT SALFT against (1) full fine-tuning ViViT and (2) the best-performing baseline. Given $N=5$, traditional significance ($p<0.05$) is mathematically unattainable; thus, the theoretical minimum ($p=0.0625$) is utilized to denote consistent empirical superiority across all folds. Full statistics are provided in the GitHub repo.

\subsection{Comparison of Efficiency}

In addition to predictive performance, we evaluate the efficiency of our proposed SALFT framework using hardware-agnostic metrics. This analysis compares SALFT applied to the ViViT model against the standard full fine-tuning of ViViT, focusing on the first fold of the 5-fold cross-validation. The exploratory full-model fine-tuning stage required by SALFT to determine layer selection is contained within this fold, making it a representative and fair basis for efficiency comparison. Other baselines (e.g., Random Forest, ResNet+LSTM) are excluded from this section due to their fundamentally different architectures and computational profiles, which would preclude a direct and meaningful comparison of fine-tuning efficiency.

\subsubsection{Parameter Efficiency}
To compare these two approaches, we quantify parameter efficiency by the absolute number of trainable parameters. We also report this value as a percentage of the total parameters of the base model. These metrics directly reflect storage cost (e.g., checkpoint size) and impact memory usage for optimizer states, providing key indicators of storage footprint and deployment practicality.
\subsubsection{Computational Efficiency Analysis}\label{sec:efficiency:compute}
We analyze computational efficiency by plotting the weighted $F_1$ score per epoch on the validation set against cumulative computational cost measured in Floating Point Operations (FLOPs). For each method under comparison, the cost per epoch ($Cost_{epoch}$) is estimated by measuring the FLOPs per fine-tuning step using \texttt{torch.profiler} after a warm-up phase, then multiplying by the number of steps per epoch $S$:
\begin{equation}
    Cost_{epoch} = FLOPs_{step} \times S.
\end{equation}

Here, $FLOPs_{step}$ is averaged over 10 iterations, each comprising a forward pass, backward pass, and optimizer update. We exclude the computational cost of validation inference, as it is negligible compared to fine-tuning. Validation involves only a forward pass, whereas fine-tuning requires both forward and backward passes, which are substantially more computationally intensive. Furthermore, the number of validation steps is typically smaller than the number of training steps, further reducing its relative impact. For our two-stage SALFT framework, the cumulative computational cost includes both the exploratory and selective fine-tuning stages, enabling a fair comparison with the baseline. Efficiency is illustrated by plotting weighted $F_1$ against cumulative FLOPs, where methods achieving higher weighted $F_1$ at lower FLOPs are deemed more efficient.

\subsection{Interpretability Methods Evaluation}\label{exp:perturbation}

\subsubsection{Baselines}
We evaluate the effectiveness of our proposed interpretability method on our arousal change recognition task using the fine-tuned model with SALFT from the first fold of the 5-fold cross-validation. Our approach is evaluated against three baselines:
\begin{itemize}
    \item \textbf{Rollout}~\cite{abnar-zuidema-2020-quantifying}: Vanilla attention rollout across all layers.
    \item \textbf{Grad-CAM}~\cite{selvaraju2017grad}: Grad-CAM adapted for the ViViT architecture, using the final encoder layer's frame token embeddings as the feature map, with gradients computed based on the \texttt{[CLS]} token's output. 
    \item \textbf{Grad-SAM}~\cite{10.1145/3459637.3482126}: Computes the element-wise product of attention maps and their ReLU-clipped gradients, averaged across all layers and heads; here, we adapt it specifically to the ViViT architecture.
\end{itemize}
In our evaluation, we exclude the class of LRP-based methods, such as those analyzed in~\cite{Chefer_2021_CVPR}. Their adaptation to the ViViT architecture incurs prohibitive computational costs that lead to out-of-memory errors with our available hardware. This practical challenge highlights the need for more computationally efficient interpretability techniques, which is a primary motivation for our interpretability method.

\subsubsection{Perturbation Analysis Protocol}
For the quantitative evaluation of our explanation method, we follow the general methodology for perturbation analysis presented in~\cite{Chefer_2021_CVPR}. For each sample, this analysis is performed against two distinct targets: (1) the model’s predicted class, to evaluate the fidelity of the explanation to the model's internal decision-making process; and (2) the ground-truth class, to assess the plausibility of the explanation by examining its ability to highlight genuinely class-discriminative features. The evaluation is conducted on the test set of each game dataset.

The protocol involves sequentially masking tubelet embeddings over 10 discrete steps (from 10\% to 90\%) by zeroing them out after the initial embedding layer but prior to the transformer encoder. This follows two complementary schemes:
\begin{itemize}
    \item \textbf{Positive Perturbation (Deletion):} The most relevant tubelets are masked first. A faithful explanation should identify crucial tokens, leading to a steep drop in model confidence upon their removal.
    \item \textbf{Negative Perturbation (Preservation):} The least relevant tubelets are masked first. In this case, model confidence should be largely maintained until important tokens are removed.
\end{itemize}
The primary metric is the Area Under the Curve (AUC) of the resulting confidence-vs-perturbation plot. For this analysis, the confidence is the model's output probability for the initially predicted class. A higher-fidelity explanation is thus characterized by a lower AUC for positive perturbation and a higher AUC for negative perturbation.

%%%%%%%%%%%%%%%%%%%%%%%%%%%%%%%%%%%

\section{Results and Discussion}\label{sec:experiment:results}
\subsection{Overall Performance Comparison}
As summarized in Table~\ref{tab:results_summary}, the proposed SALFT framework delivers strong performance across the nine AGAIN games. When compared to full fine-tuning, SALFT achieves numerically higher weighted F1 scores in 6 games, macro F1 in 5 games, and accuracy in 7 games, while updating only about 8\% of parameters. It attains the best scores across all three metrics in 5 games. The only case where full fine-tuning shows a modest, consistent lead is in Tiny. These results demonstrate that selectively adapting a minimal, task-relevant subset of parameters is sufficient to match or exceed the performance of updating all parameters, with no instances of degradation reaching the theoretical minimum $p$-value ($p \!=\! 0.0625$). The balanced results across weighted and macro F1 further indicate that this efficiency is achieved without compromising robustness to class imbalance, confirming SALFT as an effective and efficient adaptation strategy for video transformers.

\begin{table*}[tbp]
\centering
\scriptsize
\renewcommand{\arraystretch}{1.2}
\setlength{\tabcolsep}{3pt}
\caption{Values are percentages with 95\% confidence intervals in parentheses. Best result per game per metric is highlighted in \textcolor{ForestGreen}{\textbf{bold}}. Cases where ViViT SALFT yields empirical improvements reaching the theoretical minimum $p$-value ($0.0625$ for $N=5$, exact two-sided Wilcoxon signed-rank test) are denoted by $^{\dagger}$ (vs. ViViT Full) and $^{\ddagger}$ (vs. the best competing baseline).}
\label{tab:results_summary}
\begin{tabular}{l|ccc|ccc|ccc}
\hline
Method & Apex W-F1 & Apex M-F1 & Apex Acc & Endless W-F1 & Endless M-F1 & Endless Acc & FPS W-F1 & FPS M-F1 & FPS Acc \\
\hline
Random Forest        & 51.96 (1.16) & 43.00 (1.26) & 56.68 (0.78) & 53.91 (2.18) & 46.86 (2.32) & \textcolor{ForestGreen}{\textbf{59.43 (1.59)}} & 56.03 (2.85) & 46.38 (2.92) & 60.82 (2.45) \\
ResNet+LSTM Full     & 57.27 (1.43) & 53.22 (1.47) & 58.07 (1.35) & \textcolor{ForestGreen}{\textbf{58.27 (2.40)}} & \textcolor{ForestGreen}{\textbf{54.89 (2.79)}} & 58.47 (1.69) & 62.38 (1.79) & 57.73 (2.06) & 63.15 (1.60) \\
ResNet+LSTM SALFT  & 56.27 (1.65) & 52.23 (1.09) & 56.90 (2.04) & 56.39 (2.71) & 52.08 (3.52) & 57.75 (1.29) & 61.21 (2.23) & 55.98 (2.55) & 61.77 (2.64) \\
ViViT Full           & 58.05 (1.84) & \textcolor{ForestGreen}{\textbf{54.44 (2.23)}} & 58.20 (2.16) & 54.54 (1.26) & 50.58 (0.73) & 54.93 (1.79) & 62.37 (2.90) & 57.46 (3.10) & 63.03 (2.86) \\
ViViT SALFT(Ours)         & \textcolor{ForestGreen}{\textbf{58.45 (0.97)}} & 54.00 (1.69) & \textcolor{ForestGreen}{\textbf{59.13 (0.99)}} & 55.58 (1.90)$^{\dagger}$ & 51.49 (2.48) & 56.34 (2.04) & \textcolor{ForestGreen}{\textbf{62.75 (1.93)}} & \textcolor{ForestGreen}{\textbf{57.99 (2.20)}} & \textcolor{ForestGreen}{\textbf{63.58 (1.84)}} \\
\hline
Method & Gallery W-F1 & Gallery M-F1 & Gallery Acc & Gun W-F1 & Gun M-F1 & Gun Acc & Platform W-F1 & Platform M-F1 & Platform Acc \\
\hline
Random Forest        & 41.83 (2.34) & 25.38 (1.47) & 56.98 (1.83) & 42.18 (3.46) & 33.30 (2.35) & 51.33 (2.88) & 45.49 (2.67) & 36.84 (1.95) & 52.89 (2.62) \\
ResNet+LSTM Full     & 51.52 (1.77) & 44.63 (2.45) & 52.68 (2.48) & 47.90 (2.81) & 44.19 (3.27) & 48.41 (2.58) & 49.96 (2.70) & 45.61 (3.32) & 50.89 (2.69) \\
ResNet+LSTM SALFT  & 51.77 (1.30) & 44.27 (1.26) & 53.09 (2.44) & 45.65 (1.99) & 42.54 (1.58) & 45.90 (2.29) & 47.50 (3.59) & 43.38 (3.50) & 48.34 (3.84) \\
ViViT Full           & 55.29 (1.04) & 47.23 (1.55) & 57.86 (0.99) & 54.15 (3.41) & 50.84 (3.25) & 54.95 (2.92) & 54.37 (2.33) & 50.90 (2.60) & 54.98 (2.34) \\
ViViT SALFT(Ours)         & \textcolor{ForestGreen}{\textbf{56.69 (2.24)}} & \textcolor{ForestGreen}{\textbf{49.58 (2.74)}}$^{\dagger\ddagger}$ & \textcolor{ForestGreen}{\textbf{58.58 (2.36)}} & \textcolor{ForestGreen}{\textbf{57.27 (1.80)}}$^{\dagger\ddagger}$ & \textcolor{ForestGreen}{\textbf{54.08 (1.81)}}$^{\dagger\ddagger}$ & \textcolor{ForestGreen}{\textbf{57.78 (1.54)}}$^{\dagger\ddagger}$ & \textcolor{ForestGreen}{\textbf{56.39 (3.37)}} & \textcolor{ForestGreen}{\textbf{53.08 (3.51)}} & \textcolor{ForestGreen}{\textbf{57.02 (3.21)}}$^{\dagger\ddagger}$ \\
\hline
Method & Solid W-F1 & Solid M-F1 & Solid Acc & Tiny W-F1 & Tiny M-F1 & Tiny Acc & Topdown W-F1 & Topdown M-F1 & Topdown Acc \\
\hline
Random Forest        & 49.91 (2.87) & 38.12 (1.86) & 55.58 (2.33) & 50.98 (0.62) & 44.63 (0.78) & 54.66 (0.75) & 53.39 (2.00) & 44.62 (1.40) & 58.79 (1.76) \\
ResNet+LSTM Full     & 56.79 (2.10) & 51.40 (2.63) & 57.17 (2.07) & 49.29 (3.27) & 46.01 (4.13) & 50.08 (3.26) & \textcolor{ForestGreen}{\textbf{60.10 (2.02)}} & \textcolor{ForestGreen}{\textbf{56.08 (1.25)}} & 60.13 (2.50) \\
ResNet+LSTM SALFT  & 57.25 (2.08) & 51.89 (2.81) & 57.66 (2.01) & 49.79 (2.19) & 45.88 (2.67) & 50.62 (2.48) & 58.55 (2.55) & 53.80 (2.87) & 59.39 (1.84) \\
ViViT Full           & 59.43 (2.11) & 54.30 (1.27) & 59.80 (2.93) & \textcolor{ForestGreen}{\textbf{56.35 (1.66)}} & \textcolor{ForestGreen}{\textbf{53.39 (1.81)}} & \textcolor{ForestGreen}{\textbf{56.81 (1.50)}} & 58.91 (2.60) & 54.53 (2.18) & 59.04 (3.13) \\
ViViT SALFT(Ours)         & \textcolor{ForestGreen}{\textbf{60.58 (2.05)}} & \textcolor{ForestGreen}{\textbf{54.53 (2.03)}} & \textcolor{ForestGreen}{\textbf{61.41 (1.90)}} & 55.60 (2.59) & 52.11 (3.55) & 56.47 (2.19) & 59.74 (1.66) & 54.95 (1.87) & \textcolor{ForestGreen}{\textbf{60.38 (1.52)}} \\
\hline
\end{tabular}
\end{table*}

The baseline comparisons yield two critical insights. First, the limited effectiveness of the random forest baseline reinforces the necessity of deep spatiotemporal representation learning for this video-based task. Second, and more critically, SALFT's performance profile highlights its distinct advantages. It matches or surpasses full fine-tuning while updating only approximately 8\% of parameters, demonstrating that a direct, task-aware selection criterion can precisely identify a sufficient and minimal adaptation subspace. This efficiency and reliability stand in contrast to generic parameter-efficient methods. 

% to be modified
Specifically, LoRA failed to converge despite extensive hyperparameter tuning, consistently collapsing to a single-class prediction even when increasing the rank and adjusting optimization settings, suggesting that standard hyperparameter adjustments are insufficient. We hypothesize that this failure is partly due to the substantial domain shift from Kinetics-400 to gameplay videos, as the unique spatiotemporal dynamics and complex visual elements of gaming environments are likely not fully captured by the pretrained features. In addition, the arousal prediction task depends on both explicit action semantics and long-range spatiotemporal dynamics, making optimization more sensitive to representation quality and prone to degenerate solutions. This challenge is further amplified in video models such as ViViT, where spatial and temporal features are tightly coupled, potentially requiring more expressive updates. Under these conditions, LoRA’s uniform low-rank updates may be insufficient, as they constrain adaptation to a fixed low-dimensional subspace. In contrast, SALFT adopts a task-informed strategy that selectively enables unrestricted updates on responsive layers, allowing flexible adaptation while maintaining efficiency. Furthermore, its success on hybrid architectures such as ResNet+LSTM suggests that this layer selection strategy can generalize beyond transformer-based models.

In the first stage of our two-stage fine-tuning framework, the optimal layer is reliably identified within two epochs of full fine-tuning across all nine games. For the ViViT backbone, selection consistently converges on the first encoder layer (\texttt{encoder.layer.0}), whereas for the ResNet+LSTM architecture, it identifies the final ResNet-50 layer (\texttt{backbone.7}) as most responsive. We attribute this architectural divergence in layer selection to the nature of the task, which requires capturing subtle spatiotemporal dynamics in gameplay videos for arousal prediction. In ViViT, early encoder layers operate close to the embedding stage and directly process low-level spatiotemporal patterns, making them a natural point for adapting to gameplay dynamics. In contrast, in the ResNet+LSTM architecture, spatial features are extracted by the ResNet backbone and then aggregated temporally by the LSTM, making the final ResNet layer a key interface for spatiotemporal representation learning. This suggests that, driven by the task's demand for unified spatiotemporal modeling, SALFT inherently prioritizes layers where spatial and temporal information are most tightly integrated. The strong downstream performance of SALFT, achieving results comparable to full fine-tuning with only ~8\% of parameters, further suggests that this procedure can identify a minimal yet sufficient parameter subspace for adaptation.

\subsection{Analysis of Generalization Across Games}
The robust performance of SALFT across diverse games, achieving parity with full fine-tuning in 8 out of 9 cases, demonstrates its strong generalization. This success indicates that our task-aware selection criterion effectively identifies a broadly adaptable parameter subspace within the pre-trained backbone, which remains effective despite variations in genre, visual style, or perspective.

Two instructive exceptions merit closer examination. The game \texttt{Tiny}, with its extremely low-poly, isometric visuals and repetitive scenery, presents a stark domain shift where full fine-tuning retains a marginal advantage. This aligns with findings that models pre-trained on real-world imagery can struggle with simplified visual styles~\cite{jung2021improving}, and suggests that adapting to such sparse styles might benefit from broader parameter updates. In contrast, for \texttt{Endless}, which features pixel-art graphics and a highly repetitive background, the full fine-tuning of the ResNet+LSTM architecture achieves the strongest performance among all methods. This indicates that the local feature inductive bias of convolutional networks may align better with the structured patterns of pixel art for this task.

Overall, these results highlight that a targeted, task-informed adaptation strategy can efficiently overcome domain shifts caused by visual diversity. They also reveal that the choice of backbone architecture interacts with specific visual styles. The primary remaining challenge for vision-only models, however, lies in capturing emotional cues that are inherently non-visual. As noted in affective computing research, the core difficulty is in establishing a robust association between visual features and abstract emotional categories, especially when the primary triggers are non-visual~\cite{10.1145/3664647.3680875}.

\subsection{Efficiency Comparison}
\subsubsection{Parameter Efficiency}
As shown in Table~\ref{tab:efficiency}, our Second-Stage Selective Fine-Tuning updates fewer than 8\% of model parameters, compared to 100\% in full fine-tuning. This substantial reduction significantly lowers memory and storage costs. Despite the smaller parameter set, our method achieves comparable task performance across 5-fold cross-validation (Table~\ref{tab:results_summary}), indicating that selectively updating a well-chosen layer enables more efficient and effective adaptation.

\begin{table}[tbp]
\centering
\caption{Trainable parameters and per-step computational cost (FLOPs) of different fine-tuning methods (batch size of 4). Lower values (↓) indicate better efficiency.}
\label{tab:efficiency}
\resizebox{\columnwidth}{!}{%
\begin{tabular}{lcc}
\toprule
\textbf{Method} & \textbf{Trainable Params} ↓ & \textbf{FLOPs per Step (G)} ↓ \\
\midrule
Full fine-tuning & 88.65M (100\%) & 6402.90 \\
Second-Stage Selective Fine-Tuning & 7.09M (7.998\%) & 4449.00 \\
\bottomrule
\end{tabular}%
}
\end{table}

\subsubsection{Computational Efficiency}

Table~\ref{tab:efficiency} also shows the FLOPs per fine-tuning step, measured following the procedure described in Section~\ref{sec:efficiency:compute}, with a batch size of 4. FLOPs are reported in giga-FLOPs (1G = $10^9$ operations), providing a hardware-independent measure of computational cost. Our method requires only 4449.00 GFLOPs per step, reducing computational cost by 30.52\% compared to full fine-tuning (6402.90 GFLOPs). This reduction aligns with the smaller proportion of updated parameters, confirming the efficiency of our approach.

\begin{figure}[tbp]
    \centering
    \includegraphics[width=0.9\columnwidth]{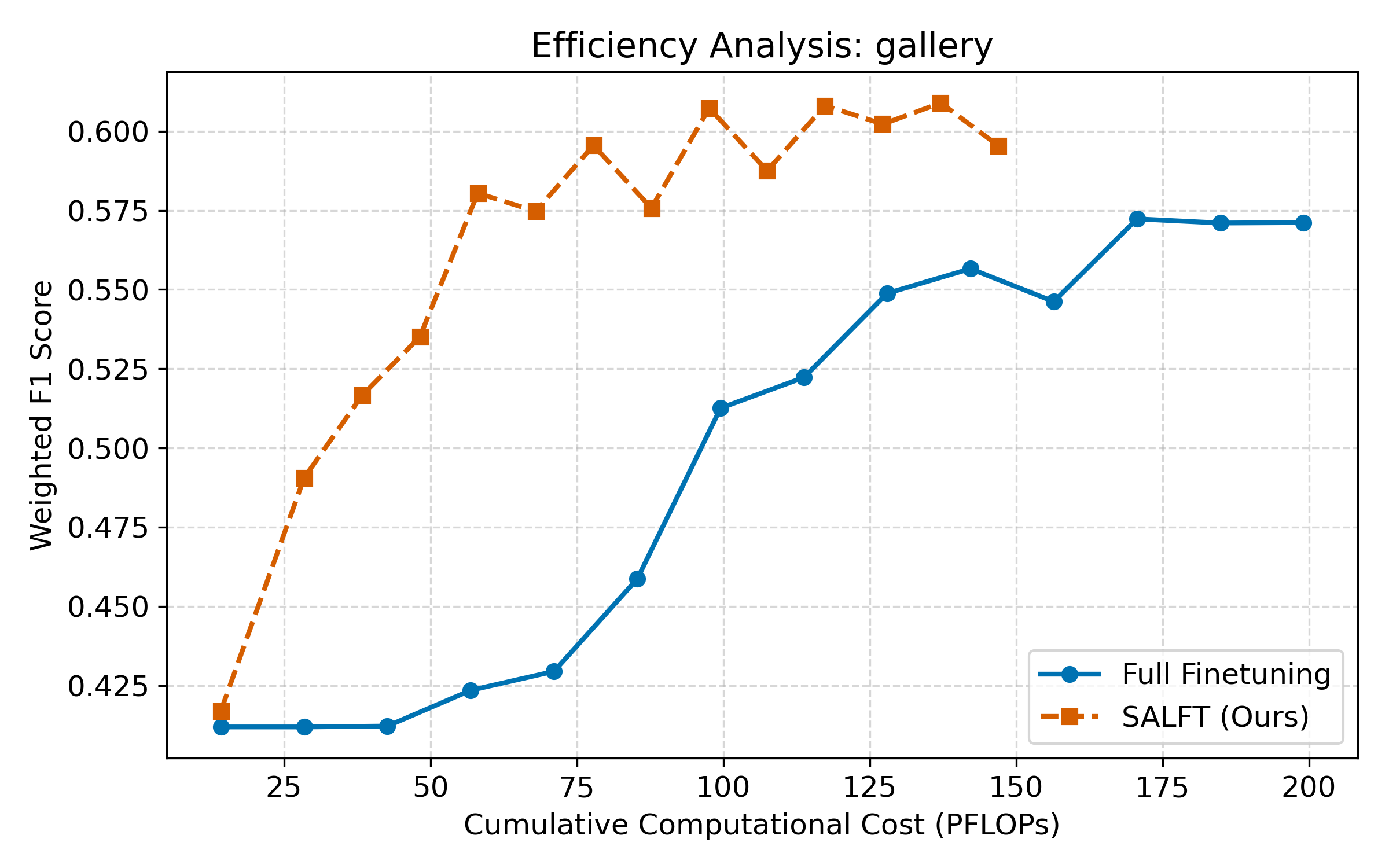}
    \caption{Weighted $F_1$ against cumulative fine-tuning FLOPs for \texttt{gallery} as a representative case (full fine-tuning vs. SALFT). Analysis is based on the first fold of 5-fold cross-validation. Complete results for all nine games are provided in the GitHub repo.}
    \label{fig:f1_flops}
\end{figure}

Figure~\ref{fig:f1_flops} plots weighted $F_1$ against cumulative fine-tuning FLOPs for the first fold, using \texttt{gallery} as a representative example (see the GitHub repo for the other eight games). The plot illustrates the trade-off between effectiveness and computational cost. Despite comprising two fine-tuning stages, our method achieves its final performance at a lower cumulative computational cost. In the exploratory first stage, training is limited to only 2 epochs to determine the layer selection, keeping this initial cost manageable. In the selective second stage, only the chosen subset of layers is fine-tuned, substantially reducing the FLOPs per epoch relative to full fine-tuning. As observed in Figure~\ref{fig:f1_flops}, SALFT typically reaches its performance plateau with less overall computation. This efficiency gain, combined with the fixed layer selection that can be reused across folds (in our 5-fold CV) or for similar tasks, accelerates model development and iteration while maintaining deployment compatibility since the final model architecture is unchanged.

\subsection{Interpretability Analysis}\label{sec:additional_analyses:interpretation}
\begin{figure*}[t]
\centering
\includegraphics[width=0.8\linewidth]{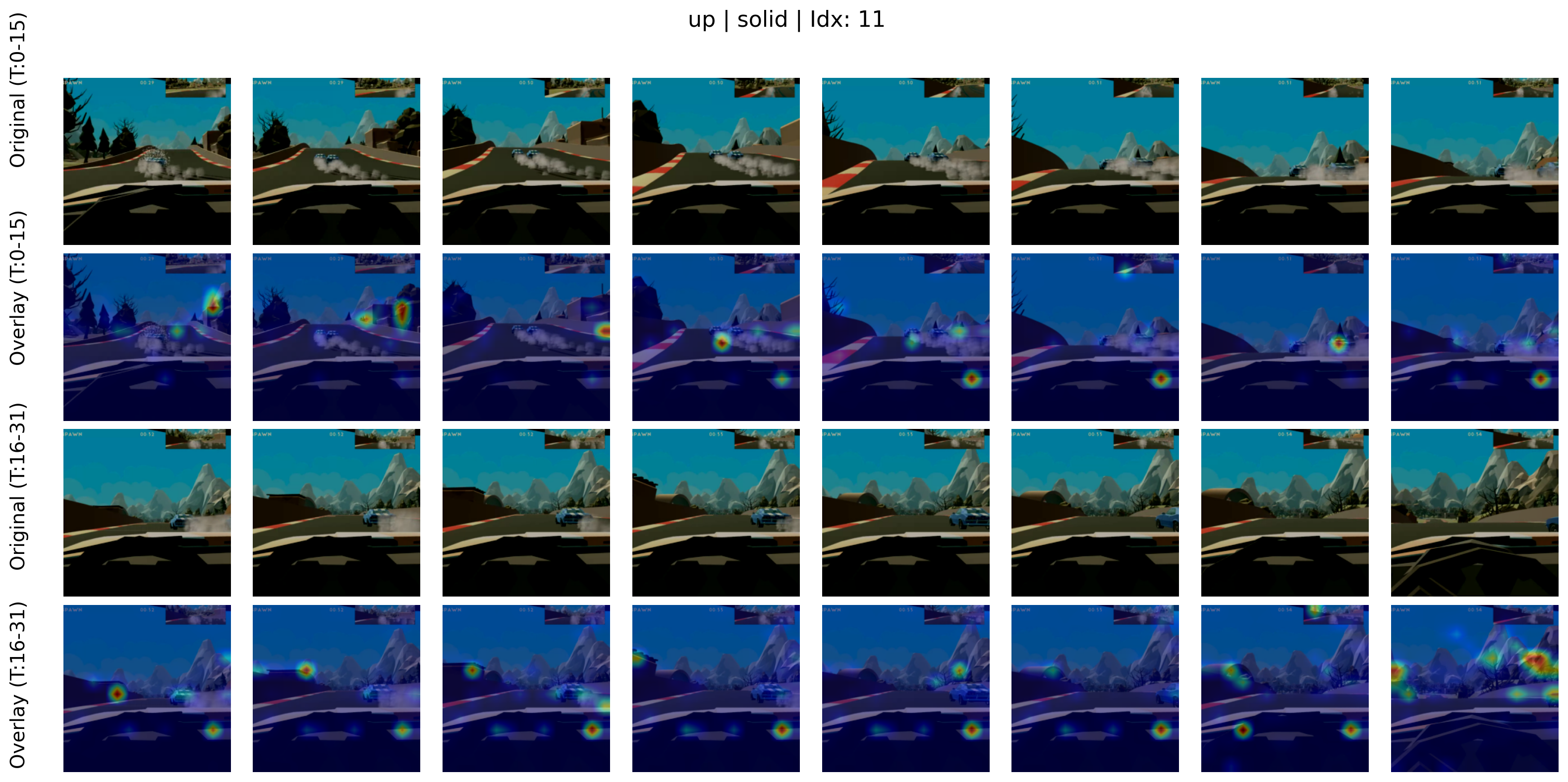}
\vspace{2mm}
\caption{Interpretation example for arousal-increasing class in \texttt{solid} (SALFT-fine-tuned model, first fold).}
\label{fig:interpretation_up}
\end{figure*}

\begin{figure*}[t]
\centering
\includegraphics[width=0.8\linewidth]{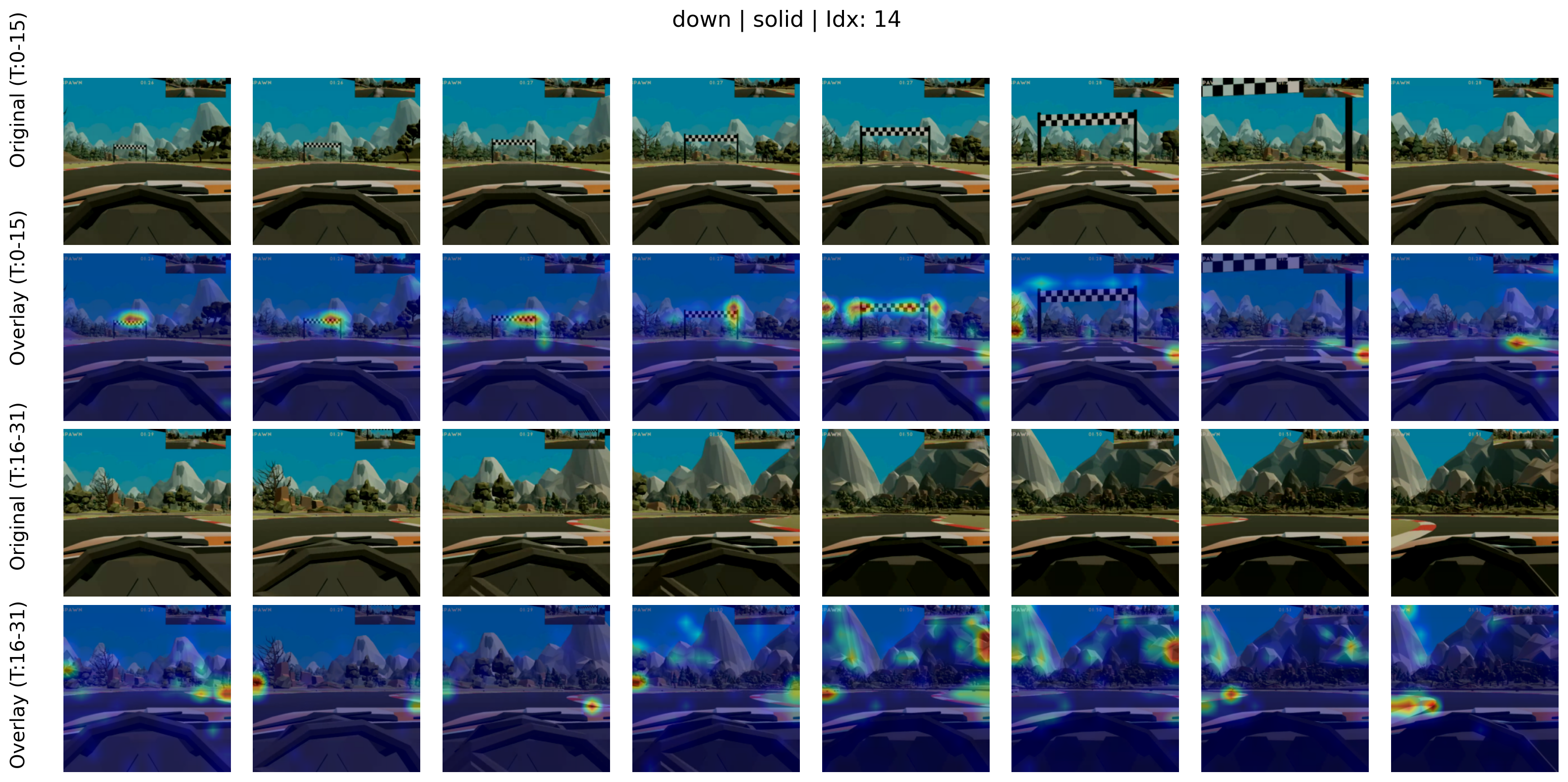}
\vspace{2mm}
\caption{Interpretation example for arousal-decreasing class in \texttt{solid} (SALFT-fine-tuned model, first fold).}
\label{fig:interpretation_down}
\end{figure*}

To understand how the fine-tuned model with SALFT (first fold of 5-fold cross-validation) responds to gameplay videos, we perform an interpretability analysis using the method described in Section~\ref{method:interpretation}. We use samples from \texttt{Solid} as a representative case due to its clear and bright visuals, which facilitate effective visualization. We selected video segments correctly classified by the model according to ground-truth labels. Figure~\ref{fig:interpretation_up} and Figure~\ref{fig:interpretation_down} show simplified examples of heatmaps overlaid on frames from \texttt{Solid}, illustrating attention patterns for arousal-increasing and arousal-decreasing segments. Warmer colors denote regions of higher relevance, indicating where the model focuses when predicting player arousal changes. Full visualizations, including segments with unchanged arousal and examples from other games, are available in the GitHub repo.

We observed that for arousal-increasing segments (Fig.~\ref{fig:interpretation_up}), the model placed greater attention on semantically meaningful regions in the latter video frames, such as enemy characters, sudden movements, bright flashes, or other salient visual events. This suggests that the model associates arousal increases with contextually relevant and potentially exciting stimuli. The model appears to exhibit a degree of semantic understanding, prioritizing emotionally salient elements over superficial cues such as static heads-up display components or background textures.

In contrast, during arousal-decreasing or arousal-unchanging segments (Fig.~\ref{fig:interpretation_down}), when the gameplay lacks dynamic or emotionally salient events, the model's attention was concentrated on visually calm or repetitive regions, such as static background elements, idle states, non-hostile navigation, or low-activity scenes. This indicates that in the absence of strong emotional cues, the model defaults to focusing on static elements of the scene, reflecting its sensitivity to the overall emotional context across different affective states.

Attention heatmaps from the SALFT-fine-tuned ViViT model exhibited game-specific patterns, demonstrating the model's ability to dynamically adjust focus according to each game's visual and gameplay context. In racing games, attention is targeted at the road ahead and potential collision zones. In shooter games, focus is centered on the crosshair, weapon barrel, enemies, and shooting-related effects. In platformers, attention is concentrated on the player character, enemies, and interactive elements essential for gameplay. Such adaptability reflects the model's sensitivity to salient visual cues and suggests robustness in handling diverse gameplay scenarios. A comparison with the full fine-tuned model's attention patterns is provided in the GitHub repo.

We further conducted a perturbation-based evaluation to quantitatively assess the attribution results, as described in Section~\ref{exp:perturbation}. Using the SALFT-fine-tuned model from the first fold, we computed AUC scores on the corresponding test set of the \texttt{solid} game as a representative example (Table~\ref{tab:auc_summary}). The proposed interpretability method performs notably well, achieving favorable AUC scores under both positive and negative perturbations, indicating that the highlighted regions are critical to the model’s decisions. Across the nine games (full results provided in the GitHub repo), our method outperforms all other baselines. These quantitative findings, based on the first-fold evaluation, corroborate our visual analysis, confirming the high faithfulness of the generated explanations.

\begingroup
\renewcommand{\textbf}[1]{\textcolor{ForestGreen}{\bfseries #1}}
\begin{table}[htbp]
\centering
\caption{Positive and negative perturbation AUC results (shown as percentages) for the predicted and target classes on the \texttt{solid} test set. Lower (↓) is better for positive perturbation and higher (↑) for negative. Best results are in \textbf{bold}.}
\label{tab:auc_summary}
\resizebox{\columnwidth}{!}{
\begin{tabular}{llcccc}
\toprule
Type & Class & Rollout & Grad-CAM & Grad-SAM & Our Method \\
\midrule
\multirow{2}{*}{Positive ↓} 
 & Predicted & 46.96 & 47.47 & 49.82 & \textbf{24.01} \\
 & Target    & 45.81 & 44.54 & 48.69 & \textbf{23.13} \\
\midrule
\multirow{2}{*}{Negative ↑} 
 & Predicted & 86.86 & 87.91 & 84.67 & \textbf{93.91} \\
 & Target    & 80.75 & 84.64 & 78.65 & \textbf{87.72} \\
\bottomrule
\end{tabular}}
\end{table}
\endgroup

\subsection{Limitations}
First, the generalizability of our layer selection strategy requires further validation across more diverse games and domains, such as those represented in recently released multimodal affective game corpora \cite{10934747,barthet2024gamevibe}. Second, as a vision-only method, its performance is inherently limited in visually sparse environments, motivating the future integration of multimodal data—an opportunity enabled by the same emerging datasets. Third, to isolate the effects of our layer selection strategy, we fixed the threshold for defining arousal changes; consequently, a full sensitivity analysis of this threshold's impact on performance is left for future work. Fourth, our focus on a single fine-tuned layer may not capture more complex feature interactions, and future work could explore multi-layer or dynamic tuning strategies.
Finally, while SALFT ViViT consistently matches or numerically exceeds both full fine-tuning ViViT and the best-performing baseline across all games, only a subset of these improvements achieves the theoretical minimum $p$-value, an inherent constraint of the 5-fold cross-validation setup.

%%%%%%%%%%%%%%%%%%%%%%%%%%%%%%%%%%%

\section{Conclusions}
This paper introduced Selective Adaptive Layer Fine-Tuning (SALFT), a parameter-efficient framework for recognizing player arousal from gameplay videos. By fine-tuning only the most responsive transformer layers identified via an L2-norm criterion, SALFT reduces trainable parameters by approximately 92\% (from 88.65M to 7.09M) and training FLOPs by 30.5\%. As validated by exact Wilcoxon signed-rank tests, SALFT ViViT maintains performance statistically comparable to both full fine-tuning ViViT and the best-performing baseline across all nine games, succeeding where the generic PEFT method, LoRA, failed to converge. Interpretability analyses reveal that SALFT dynamically attends to semantically meaningful game elements across different genres. These results position SALFT as a practical and robust solution for resource-constrained affective computing in gaming, offering an effective balance between performance, efficiency, and interpretability.

\bibliographystyle{IEEEtran}
\bibliography{refs}
% \section*{Appendix--}
\end{document}